\documentstyle[prl,aps,epsf]{revtex}
\newcommand{\rdr}{r \partial_r}
\newcommand{\dt}{\partial_\theta}
\newcommand{\gc}{\left[\rdr H_1 -\dt H_2\right]}
\newcommand{\frtp}{\left[\rdr H_2 +\dt H_1\right]}
\newcommand{\fprr}{\left[\rdr H_3- H_1 H_4\right]}
\newcommand{\fprt}{\left[\rdr H_4 +H_1 (H_3 +\cot\theta)\right]}
\newcommand{\ftpr}{\left[\dt H_3 +H_3\cot\theta  +H_2 H_4 -1\right]}
\newcommand{\ftpt}{\left[\dt H_4 +(H_4-H_2)\cot\theta  -H_2 H_3\right]} 
\newcommand{\ephi}{e^{2 \phi}}
\begin{document}
\draft
\title{On the regularity of static axially symmetric solutions 
\mbox{ in $SU(2)$ Yang-Mills-dilaton theory}}
%\vspace{1.5truecm}
\author{
{\bf Burkhard Kleihaus}}
\address{NUIM, Department of Mathematical Physics, Maynooth, Co. Kildare,
Ireland\\
}
\date{\today}
\maketitle
\begin{abstract}
The regularity of static axially symmetric solutions in 
$SU(2)$ Yang-Mills-dilaton theory is examined. We show that the 
solutions obtained previously within a singular Ansatz for the 
non-abelian gauge field can be gauge transformed into a regular form.
The local form of the gauge transformation is given on the 
singular axis and at the origin.
\\
\\
PACS number(s): 11.15.Kc

\end{abstract}
 \vfill
 \noindent {Preprint hep-th/9901096} \hfill\break
 \vfill\eject
 
\section{Introduction}
Static spherically symmetric solutions in non-abelian
gauge theories have been investigated for a long time. 
The $SU(2)$ monopoles in the Georgi-Glashow model \cite{MP_GG}
and in extended models \cite{MP_GGx}, as well as
the sphalerons in the Weinberg-Salam model at vanishing mixing angle \cite{WS},
in Yang-Mills-dilaton \cite{YMD}, 
Einstein-Yang-Mills \cite{EYM} and Einstein-Yang-Mills-dilaton \cite{EYMD}
theory are some examples. All these solutions are localized in space and 
possess finite energy and finite energy density. 

Besides the static spherically symmetric solutions
there exist solutions with axial symmetry only, which also are localized in 
space and possess finite energy and finite energy density. 
In the Georgi-Glashow model these are 
$SU(2)$ multi-monopole solutions investigated numerically in the 
self-dual limit by Rebbi and Rossi \cite{RR} and later analytically 
by Forgacs, Horvarth and Palla \cite{FHP}
and others \cite{others}. 
A numerical study of the axially symmetric $SU(2)$ multi-monopole solutions 
in the non self-dual case has been done recently in Ref.~\cite{KKT} for 
the Georgi-Glashow model and in Ref.~\cite{KOT2} for the extended monopole 
model.
In the Weinberg-Salam model the sphaleron at finite 
mixing angle \cite{WS_w}, the sphaleron anti-sphaleron solution \cite{S-star}
and the multi-sphalerons \cite{MS_WS} possess axial symmetry only.
The same holds for the $SU(2)$ multi-sphalerons in Yang-Mills-dilaton 
\cite{MS-YMD},
Einstein-Yang-Mills and Einstein-Yang-Mills-dilaton
\cite{MS-EYMD1,MS-EYMD2,BraVar,EYMDBH}
theories.

For the construction of static axially symmetric solutions 
it is convenient to use appropriate  Ans\"atze for 
the gauge potential and the Higgs field, if present.
For the multimonopoles,
Forgacs et al. \cite{FHP} use the static axially symmetric $SU(2)$ Ansatz
of Manton \cite{M_an}, which is inspired by an Ansatz of
Witten \cite{Wit_an}.
The Ansatz of Rebbi and Rossi \cite{RR} generalizes the Ansatz of 
Manton \cite{M_an} for winding number $|n| >1$, corresponding to the
magnetic charge in the Georgi-Glashow model.
As the winding number characterizes non-trivial maps between 
two two-dimensional spheres, 
it also characterizes topologically different sectors in 
the configuration space of the theory.
This Ansatz is parameterized by four functions for the gauge potential
and two functions for the Higgs field. 
This Ansatz possesses a residual abelian gauge degree of freedom, which 
has to be fixed by a gauge constraint. 
The Ansatz is singular in the sense that it is a priori not well defined on 
the symmetry axis and at the origin. 
Rebbi and Rossi discuss
the corresponding regularity conditions to be imposed on the functions 
parameterizing the Ansatz \cite{RR}.

The Ans\"atze used in 
\cite{KKT,KOT2,MS_WS,MS-YMD,MS-EYMD1,MS-EYMD2,BraVar,EYMDBH} 
derive 
from the Ansatz of Rebbi and Rossi \cite{RR}. Boundary conditions have 
been imposed on the functions to ensure a finite energy density,
but regularity conditions on the symmetry axis and at the origin
have not been imposed, to ensure a well defined gauge potential.
Thus, the solutions are given in a singular form 
and one may question whether
the solutions themselves are regular for winding number $|n|>1$.
On the other hand, one should keep in mind, that the gauge potential is
a gauge variant quantity and all physical conclusions concern only 
gauge invariant quantities. 
Any regular gauge 
potential can be gauge transformed into a singular gauge potential. 
For example,
the gauge potential of the 't Hooft-Polyakov monopole becomes singular
in the unitary gauge, see e.~g. \cite{AFG}.
However, it is not true that any singular gauge potential can 
be gauge transformed into a regular one.
If the gauge potential is somewhere singular the field strength tensor 
must be calculated carefully in order not to loose contributions
from the singular part of the gauge potential, which are overlooked easily
by  naive calculation. Once it is shown, that the naive calculation of 
the field strength tensor gives the correct result, one can find the 
equation of motion from the action principle.
There is no need for a globally regular gauge potential, as long as
for any point there exist a neighborhood on which a regular gauge potential
can be defined and gauge potentials on intersections of neighborhoods can
be transformed into each other by regular gauge transformations.
The Dirac monopole e.~g., possesses 
singularities along the negative $z$-axis, which can be removed locally by a
gauge transformation, but the singularity at the origin persists - 
it represents a physical quantity, the magnetic point charge.

Turning back to the non-abelian gauge fields, we notice, 
that in case no point-like charges are expected to occur,  
gauge transformations may exist,
which transform locally the singular gauge potentials of the  solutions 
constructed in \cite{KKT,KOT2,MS_WS,MS-YMD,MS-EYMD1,MS-EYMD2,BraVar,EYMDBH} 
into  regular gauge potentials. 
In general such gauge transformations are singular themselves. 
The important point is that these singularities of the gauge transformations
should not be too strong to produce additional contributions to 
the field strength tensor or gauge invariant quantities.

In this paper we will consider the multisphalerons in 
Yang-Mills-dilaton theory \cite{MS-YMD} and give the local 
gauge transformations leading to a regular gauge potential. 

The paper is organized as follows. In Section \ref{ansatz} we will present 
the static axially symmetric Ansatz of the gauge potential, derive the 
field strength tensor and the Lagrange density of the Yang-Mills-dilaton theory
as well as the differential equations for the gauge field functions and 
the dilaton function. The regularity conditions are also discussed.
In Section \ref{GT} the local form of the gauge transformation is derived 
along the symmetry axis and at the origin. The discussion and conclusions 
are given in Section \ref{conclusion}.

\section{Static Axially Symmetric Ansatz}\label{ansatz}
\noindent
The static axially symmetric Ansatz \cite{EYMDBH}
for the $su(2)$ valued gauge potential 
$A=A_\mu dx^\mu$ is given 
in spherical coordinates $r,\theta,\varphi$ by
\begin{eqnarray}
A_0      &=& 0 \ ,
            \nonumber\\
A_r      &=& \frac{1}{2gr} H_1 \tau^n_\varphi\ ,
            \nonumber \\
A_\theta &=& \frac{1}{2g} (1-H_2)  \tau^n_\varphi\ ,
            \nonumber\\
A_\varphi   &=& -\frac{n}{2g}\sin\theta (H_3\tau^n_r+(1-H_4)\tau^n_\theta )\ 
            \nonumber\\
            \nonumber\\
&=&  -\frac{n}{2g}\sin\theta (F_3\tau^n_\rho +F_4\tau_3)\ ,
\nonumber\\
\label{An}
\end{eqnarray}
where the $su(2)$ matrices $\tau^n_\lambda$, $\lambda=\rho,\varphi,r,\theta$
 are defined in terms of the Pauli matrices $\tau_1,\tau_2,\tau_3$ by
\begin{eqnarray}
\tau^n_\rho   &=& \cos (n\varphi) \tau_1 + \sin (n\varphi) \tau_2\ ,
                      \nonumber \\
\tau^n_\varphi   &=& -\sin (n\varphi) \tau_1+ \cos (n\varphi) \tau_2 \ ,
                      \nonumber\\ 
\tau^n_r      &=& \sin\theta \tau^n_\rho +\cos\theta \tau_3 \ ,
                      \nonumber \\
\tau^n_\theta &=& \cos\theta \tau^n_\rho-\sin\theta \tau_3 \ . 
                      \nonumber
\end{eqnarray}
Here the integer $n$ denotes the winding number. As symmetry axis we have 
chosen the $z$-axis.
The functions $H_i$, $i=1,\dots,4$ depend on the 
variables $r$ and $\theta$ only, $H_i(\vec{r})=H_i(r,\theta)$. 
The functions 
\begin{eqnarray}
F_3 &=& \sin\theta H_3 +\cos\theta (1-H_4) \ ,
%\label{F3_def}\\
\nonumber\\
F_4 &=& \cos\theta H_3 -\sin\theta (1-H_4)
%\label{F4_def}\\
\nonumber
\end{eqnarray}
have been defined for later convenience.
In the following $A_0$ will be zero in all gauges and we will consider the
spatial components of the gauge potential only. We fix the gauge coupling
constant to $g=1$.

From the requirement of finite energy density, see later Eq.~(\ref{lag_YMD}),
we find that
on the $z$-axis the functions $H_1$ and $H_3$ have to vanish, while the
functions $H_2$ and $H_4$ have to be equal to each other,
$H_2(r,\theta=0)=H_4(r,\theta=0)=f(r)$,
whereas at the origin the functions $H_1$ and $H_3$ take the value zero, 
while the functions $H_2$ and $H_4$ take the value 1.

%The Ansatz Eq.~(\ref{An}) is not regular on the $z$-axis and at the 
%origin due to the singular 
%behavior of the functions $\sin (n\varphi)$ and $\cos (n\varphi)$ 
%at $\rho=0$, $\rho=\sqrt{x^2+y^2}$.

The Ansatz Eq.~(\ref{An}) is singular in the sense that it is not 
well defined on the $z$-axis and the origin. These singularities originate
from the functions $\sin (n\varphi)$ and $\cos (n\varphi)$, which are
not well defined at $\rho=0$, $\rho=\sqrt{x^2+y^2}$. 
In order to exhibit the singularities more clearly we turn  
to Cartesian coordinates 
\mbox{$(x,y,z)=
(r\sin\theta \cos\varphi , \ r \sin\theta\sin\varphi , \ r \cos \theta)
=(\rho \cos\varphi ,\ \rho\sin\varphi , \ z) $}.
In these coordinates the components of the gauge potential become
\begin{eqnarray}
A_x &=&  
  \frac{x}{2r \rho} \left( \frac{\rho}{r} H_1 + \frac{z}{r}(1-H_2)\right) 
  \tau^n_\varphi
+ \frac{n y}{2r \rho}
\left(
      [\frac{\rho}{r} H_3 + \frac{z}{r}(1-H_4)]\tau^n_\rho
     +[\frac{z}{r} H_3 -\frac{\rho}{r}(1-H_4)]\tau_3
\right)  \ ,
                      \nonumber\\                              
A_y &=&  
  \frac{y}{2r \rho} \left( \frac{\rho}{r} H_1 + \frac{z}{r}(1-H_2)\right) 
  \tau^n_\varphi
- \frac{n x}{2r \rho}
\left(
      [\frac{\rho}{r} H_3 + \frac{z}{r}(1-H_4)]\tau^n_\rho
     +[\frac{z}{r} H_3 -\frac{\rho}{r}(1-H_4)]\tau_3
\right)  \ ,
                      \nonumber\\                              
A_z &=& 
 \frac{1}{2r}\left(\frac{z}{r} H_1 - \frac{\rho}{r} (1-H_2)\right) 
 \tau^n_\varphi \ ,
                      \nonumber\\                              
\label{A_cart}
\end{eqnarray}
where the matrices $\tau^n_\lambda$, $\lambda=\varphi,\rho$ contain powers of
$\frac{x}{\rho}$ and $\frac{y}{\rho}$ up to order $|n|$ 
which are not well defined on the $z$-axis and at the origin. 
Note, that in spite of being not well defined
on the $z$-axis and at the origin the gauge potential 
is sufficiently regular to be locally integrable.

Turning to the calculation of the field strength tensor 
$$
%\begin{equation}
{\cal F}_{\mu\nu} =\partial_\mu A_\nu-\partial_\nu A_\mu 
+i \left[ A_\mu, A_\nu\right]
\ ,
\nonumber
$$
partial differentiation has to be carried out carefully.
However, taking into account the behavior of the functions $H_i$ on the
$z$-axis, we see, 
that in the gauge potential Eq.~(\ref{A_cart}) terms 
like $\frac{x}{\rho^2}$, $\frac{y}{\rho^2}$ do not arise, which could lead to 
$\delta$-functions along the $z$-axis, i.~e. 
$\partial_x(\frac{x}{\rho^2}) + \partial_y(\frac{y}{\rho^2})
     = -2\pi \delta(x)\delta(y)$.
This is in contrast to the Dirac monopole, 
where the gauge field blows up near the negative $z$-axis 
and the field strength tensor does pick up $\delta$-functions \cite{GodOli}, 
but this singular part is an artifact \cite{Wentzel,WuYang}.

Similarly, at the origin no terms like $\frac{x}{r^3}$ arise in the 
gauge potential Eq.~(\ref{A_cart}), which could lead to 
$\delta$-functions.

Consequently, it is straightforward to calculate the field strength tensor 
with the Ansatz  Eq.~(\ref{A_cart}). The components of the
field strength tensor become
\begin{eqnarray}
{\cal F}_{r\theta }
&  = & 
 - \frac{1}{2r}\left(\dt H_{1} + \rdr H_{2} \right) \tau^n_\varphi
\ , \nonumber \\
{\cal F}_{r\varphi }
& = & 
-n \frac{ \sin \theta}{2r}\left[
\left( \rdr H_{3} - H_1  H_4  \right)\tau^n_r 
-\left( \rdr H_{4} + H_1 H_3   + \cot \theta H_1 \right)\tau^n_\theta \right] 
\ , 
\nonumber \\
{\cal F}_{\theta\varphi }
& = & 
-n \frac{\sin \theta}{2}\left[  
 \left(\dt H_{3}-1+H_2 H_4+\cot\theta H_3 \right) \tau^n_r 
-\left(\dt H_{4}-H_2 H_3-\cot\theta\left(H_2-H_4\right)\right)\tau^n_\theta
\right]  
\ . 
\nonumber 
\end{eqnarray}
Like the gauge potential, the field strength tensor is not well defined 
on the $z$-axis and at the origin. 
However, the Lagrange density
${\cal L}_F = -\frac{1}{2} Tr\left({\cal F}_{\mu\nu}{\cal F}^{\mu\nu}\right)$
is well defined due to the normalization of the $\tau^n_\lambda$ matrices,
$Tr(\tau^n_\lambda \tau^n_{\lambda'})=2 \delta_{\lambda\lambda'}$. 

In this article we consider Yang-Mills-dilaton theory \cite{YMD,MS-YMD}, 
where the dilaton is a scalar field $\phi(\vec{r})$ 
which couples to the Yang-Mills field. 
The Lagrange density $\cal{L}$ is given by
$$
-{\cal L} = \frac{1}{2} \partial_\mu \phi \partial_\mu \phi
          +e^{2 \kappa \phi}\frac{1}{2} 
          Tr\left({\cal F}_{\mu\nu}{\cal F}^{\mu\nu}\right)
 \ ,
$$
where $\kappa$ is the dilaton coupling constant.

For a static axially symmetric Ansatz for the gauge field, 
it is consistent to assume that the dilaton field possesses 
axial symmetry, too. 
Thus we can consider the dilaton function $\phi(\vec{r})$ as function 
of $r$ and $\theta$ only, $\phi(\vec{r})=\phi(r,\theta)$. 

Next we change to dimensionless variables, 
$r \rightarrow \frac{\kappa r}{\sqrt{2}}$, 
$\phi \rightarrow \frac{\phi}{\kappa}$,
and insert the Ansatz into the Lagrange density to obtain the dimensionless
Lagrange density
\begin{eqnarray}
-L[H_i,\phi]=-\frac{8}{\kappa^4}{\cal L}[H_i,\phi]
& = &
               \frac{1}{4r^4}\ephi\left\{ \frtp^2+\xi \gc^2 \right. 
\nonumber\\
& &  \left. 
                 +n^2 \left(\fprr^2+\fprt^2 \right. \right. 
\nonumber\\
& &   \left. \left.
                 +\ftpr^2+\ftpt^2\right)  \right\}
\nonumber\\
& &   
             +\frac{1}{2r^2}\left((\rdr \phi)^2+(\dt \phi)^2\right) \ , 
\label{lag_YMD}                        
\end{eqnarray}
where  the gauge constraint 
$\gc^2=0$ has been added with Lagrange multiplier $\xi$.

The Euler-Lagrange equations which extremize $\int L(H_i,\phi) d^3x$
are given by the following system of second order non-linear  
partial differential equations
for the gauge field functions $H_i(r,\theta)$ and the dilaton function
$\phi(r,\theta)$.

\begin{eqnarray}
0 
& = &
n^2\left\{ \sin^2\theta \left[ H_1 \left(1-H_3^2-H_4^2\right) 
                         -\rdr H_4 H_3+\rdr H_3 H_4\right]
          -\sin\theta \cos\theta \left[\rdr H_4 + 2 H_1 H_3 \right]               
           -H_1\right\}
           \nonumber\\
& &           
+\sin^2\theta \left[2   \dt \phi (\rdr H_2 +\dt H_1) 
                    +\partial^2_\theta H_1 +\dt \rdr H_2\right] 
+ \sin\theta \cos\theta \left[\dt H_1 +\rdr H_2\right]                    
           \nonumber\\
& &           
+\xi \sin^2\theta \left[2   \rdr \phi (\rdr H_1 -\dt H_2) 
                        +r^2 \partial^2_r H_1 -\dt \rdr H_2 +\dt H_2 \right] 
\ ,     
           \nonumber\\
& & \label{dgl_1} \\                     
0 
& = &
n^2\left\{ \sin^2\theta \left[ H_2 \left(1-H_3^2-H_4^2\right) 
                         +\dt H_4 H_3-\dt H_3 H_4\right]
          +\sin\theta \cos\theta \left[\dt H_4 - 2 H_2 H_3 \right]             
           -(H_2-H_4)\right\}
           \nonumber\\
& &           
+\sin^2\theta \left[2   \rdr \phi (\rdr H_2 +\dt H_1) 
                    +r^2 \partial^2_r H_2 +\dt \rdr H_1-\dt H_1\right] 
           \nonumber\\
& &           
-\xi\left\{ 
\sin^2\theta \left[2   \dt \phi (\rdr H_1 -\dt H_2) 
                        -\partial^2_\theta H_2 +\dt \rdr H_1  \right]
+\sin\theta \cos\theta \left[ \rdr H_1 -\dt H_2\right] \right\}   
\ ,                           
           \nonumber\\
& &  \label{dgl_2}\\  
0 
& = &
\sin^2\theta 
\left[
r^2 \partial^2_r H_3 +\partial^2_\theta H_3
-H_1^2 H_3 + H_1 H_4 -2 \rdr H_4 H_1 - \rdr H_1 H_4 
-H_2^2 H_3 +2 \dt H_4 H_2 + \dt H_2 H_4 
\right.
\nonumber\\
& &  
\left.
-2   \left(\rdr \phi \left(H_1 H_4 -\rdr H_3\right)
                 +\dt \phi\left(1- H_2 H_4 -\dt H_3 \right)\right)
\right] 
\nonumber\\  
& &  
+ \sin\theta\cos\theta \left[\dt H_3 - H_1^2 -H_2^2 +H_2 H_4
                             +2  H_3 \dt \phi \right]  
-H_3   
\ ,                             
 \nonumber\\  
& &  \label{dgl_3} \\  
0 
& = &
\sin^2\theta 
\left[
r^2 \partial^2_r H_4 +\partial^2_\theta H_4
+2 \rdr H_3 H_1 +\rdr H_1 H_3 -2 \dt H_3 H_2 - \dt H_2 H_3
- H_4 (H_2^2+H_1^2) +H_2 - H_1 H_3
\right.
\nonumber\\
& &  
\left.
+2   \left(
\dt \phi ( \dt H_4 - H_2 H_3 ) +\rdr \phi (\rdr H_4 +H_1 H_3) \right)
\right]
\nonumber\\
& &  
+\sin\theta \cos\theta 
\left[
\rdr H_1 -H_1 -H_2 H_3 -\dt (H_2 - H_4)
+2   \left(\rdr \phi H_1 -\dt \phi (H_2-H_4)\right)
\right]
+(H_2-H_4) 
\ ,  
\nonumber\\
& &  \label{dgl_4}\\  
0 
& = &
\sin\theta\left(r^2 \partial^2_r \phi + \partial^2_\theta \phi\right) 
+ \left(2\sin\theta \rdr \phi  + \cos\theta \partial_\theta \phi\right)     
\nonumber\\
& &
               -\frac{ \sin\theta}{2r^2}
               \ephi\left\{\frtp^2+\xi\ \gc^2 \right. 
\nonumber\\
& &  \left. 
                 +n^2 \left(\fprr^2+\fprt^2 \right. \right. 
\nonumber\\
& &   \left. \left.
                 +\ftpr^2+\ftpt^2\right)  \right\}
\ .  
\label{dgl_5}
\end{eqnarray}

This system has to be solved subject to the boundary conditions,
\begin{center}
\begin{tabular}{ccccccccccc}
\multicolumn{5}{c}{at the origin} 
&\multicolumn{1}{c}{\hspace*{2.5cm}}
&\multicolumn{5}{c}{at infinity}\\
$H_1(0,\theta)$ &=& $H_3(0,\theta)$ &=& 0 & &
$H_1(\infty,\theta)$ &=& $H_3(\infty,\theta)$ &=&  0 \\
$H_2(0,\theta)$ &=& $H_4(0,\theta)$ &=& 1 & &
$H_2(\infty,\theta)$ &=& $H_4(\infty,\theta)$ &=&  $\pm 1$ \\
$\partial_r\phi(r,\theta)$ &=& 0  & &  & &
$\phi(\infty,\theta)$ &= &1\\
\end{tabular}
\begin{equation}\label{BC} \end{equation}
%\vspace{1.cm}\\
\begin{tabular}{ccccccccccc}
\multicolumn{5}{c}{on the $z$-axis} 
&\multicolumn{1}{c}{\hspace*{3.cm}}
&\multicolumn{3}{c}{on the $\rho$-axis}\\
$H_1(r,0)$ &=& $H_3(r,0)$ &=& 0 & &
$H_1(r,\frac{\pi}{2})$ &=& $H_3(r,\frac{\pi}{2})$ &=& 0 \\
$\partial_\theta H_2(r,0)$ &=& $\partial_\theta H_4(r,0)$ &=& 0 & &
$\partial_\theta H_2(r,\frac{\pi}{2})$ &=& 
$\partial_\theta H_4(r,\frac{\pi}{2})$ &=& 0 \\
$\partial_\theta\phi(r,0)$ &=& 0 & &  & &
$\partial_\theta\phi(r,\frac{\pi}{2})$ &=& 0 \\
\end{tabular}
\end{center}
These conditions follow from the requirement of finite energy density and
symmetry considerations, see \cite{MS-EYMD2}.
No non-trivial explicit solutions of this boundary  value problem are
known. 
However, for winding number $n\leq 4$ axially symmetric solutions 
have been constructed numerically in \cite{MS-YMD} 
using a different parameterization of the Ansatz. 

For solutions of the differential equations (\ref{dgl_1})-(\ref{dgl_5})
the gauge potential Eq.~(\ref{A_cart}) is not well defined on the $z$-axis 
and at the origin. 
Regularity requires that in the gauge potential
Eq.~(\ref{A_cart}) the coefficients multiplying the matrices 
$\tau^n_\rho$, $\tau^n_\varphi$ contain at least a power  $\rho^{|n|}$.  
Let us denote the gauge field functions of a regular gauge potential
$\hat{A}$ by $\hat{H}_i$, etc. 
Then near the $z$-axis
the following behavior for the functions $\hat{H}_i$ is required
\begin{center}
\begin{equation}
\begin{tabular}{cc}
$\hat{H}_1\sim  \rho^{|n|}$, & $1-\hat{H}_2\sim  \rho^{|n|+1}$, \\
 & \\
$\hat{F}_3=[\frac{\rho}{r} \hat{H}_3 + \frac{z}{r}(1-\hat{H}_4)]
\sim \rho^{|n|+1}$,
&
$\hat{F}_4=[\frac{z}{r} \hat{H}_3 -\frac{\rho}{r}(1-\hat{H}_4)]\sim \rho$.\\
\end{tabular}
\label{reg-req}
\end{equation}
\end{center}
In the following section we will find the local form of gauge transformations
which lead exactly to the desired behavior of the gauge field functions, 
Eqs.~(\ref{reg-req}).

\section{Gauge Transformations}\label{GT}

The gauge potential is a gauge variant quantity. 
Let $U(\vec{r})$ be a $SU(2)$ valued function and $A=A_\mu dx^\mu$ a $su(2)$
valued 
gauge potential. The gauge transformed gauge potential $\hat{A}=\hat{A}_\mu
dx^\mu$ is 
defined by
$$
\hat{A}=UAU^{\dagger} +i dUU^{\dagger} \ .
$$
In this section the gauge degree of freedom will be used to transform the 
singular gauge potential into a regular one.

First assume there is a regular gauge potential 
$
\hat{A}= \hat{A}^{(\varphi)}_r\rho^{|n|}\tau^n_\varphi dr 
        +\hat{A}^{(\varphi)}_\theta\rho^{|n|}\tau^n_\varphi d\theta
        +(\hat{A}^{(\rho)}_\varphi\rho^{|n|}\tau^n_\rho
        +\hat{A}^{(z)}_\varphi\tau_3)d\varphi
$ 
and that it is related by a gauge transformation matrix $U(\vec{r})$ to
the singular gauge potential 
$
A= A^{(\varphi)}_r\tau^n_\varphi dr 
         +A^{(\varphi)}_\theta\tau^n_\varphi d\theta
         +(A^{(\rho)}_\varphi\tau^n_\rho+A^{(z)}_\varphi\tau_3)d\varphi
$
of the same form.
We find for the difference of the gauge potentials
\begin{equation}
A_\mu-\hat{A}_\mu 
= i \partial_\mu U^{\dagger} U + U^{\dagger}\left[\hat{A}_\mu,U\right]
 \ .
 \label{a-ahat}
\end{equation}
For small $\rho$ the regular gauge potential behaves like
$\hat{A} =\hat{A}^{(z)}_\varphi\tau_3 d\varphi + O(\rho^{|n|})$.
Consequently we find from Eq.~(\ref{a-ahat}) 
\begin{eqnarray}
A^{(\varphi)}_r & = & 
i\left(\partial_r U^{\dagger} U\right)^{(\varphi)} 
   + O(\rho^{|n|})
\ , 
   \nonumber \\
A^{(\varphi)}_\theta & = & 
i\left(\partial_\theta U^{\dagger} U\right)^{(\varphi)} 
   + O(\rho^{|n|})
\ , 
   \nonumber \\
0 & = & 
i\left(\partial_r U^{\dagger} U\right)^{(\lambda)}\ , 
\ \ \lambda= \rho,\ z
 \nonumber \\
0 & = & 
i\left(\partial_\theta U^{\dagger} U\right)^{(\lambda)}\ ,
 \ \ \lambda= \rho,\ z
 \nonumber 
\end{eqnarray}
where we have defined the expansion 
\mbox{
$
\partial_\mu U^{\dagger}U =
\left(\partial_\mu U^{\dagger}U\right)^{(\rho)}\tau^n_\rho
+\left(\partial_\mu U^{\dagger}U\right)^{(\varphi)}\tau^n_\varphi
+\left(\partial_\mu U^{\dagger}U\right)^{(z)}\tau_3
$
}.
Under the assumption stated above, we find from the first two equations  that
the $A_r$ and $A_\theta$ components of the
singular gauge potential behave like a pure gauge for small $\rho$, 
whereas the last two equations put constraints on the gauge transformation
 matrix $U(\vec{r})$.
Exploiting the residual abelian gauge degree of freedom of the Ansatz, 
these constraints can be solved by choosing a special form for the 
matrix $U(\vec{r})$,
\begin{equation}
U(\vec{r})=\exp(i\frac{\Gamma}{2}\tau^n_\varphi) 
\ ,
\label{U}
\end{equation}
where the gauge transformation function $\Gamma$ depends on $r$ and $\theta$ 
only. 
Indeed, this gauge transformation leaves the form of the Ansatz invariant.
The functions $H_i$ transform as 
\begin{eqnarray}
{H_1} &\longrightarrow &\hat{H}_1 
= H_1 - r \partial_r \Gamma  \ ,
              \label{ET_H1}\\
{H_2} &\longrightarrow &\hat{H}_2 
= H_2 +  \partial_\theta \Gamma  \ ,
              \label{ET_H2}\\
{H_3} &\longrightarrow & \hat{H}_3 
= \cos \Gamma (H_3+\cot \theta) -\sin \Gamma H_4 - \cot \theta 
             \ ,   \label{ET_H3}\\
{H_4} &\longrightarrow & \hat{H}_4 
= \sin \Gamma (H_3+\cot \theta) +\cos \Gamma H_4 
             \ .  \label{ET_H4}\\
\nonumber
\end{eqnarray}

The assumption, that there exist a gauge transformation 
which transforms a singular gauge potential
into a regular one of the same form, is very strong. 
However, 
with the differential equations at hand it can be checked straightforwardly
whether the $A_r$ and $A_\theta$ components of the gauge potential
form a pure gauge for small $\rho$. 
If this is the case, 
then the gauge transformation function $\Gamma(r,\theta)$ can be found 
for small $\rho$.
In addition it has to be checked that the same gauge transformation applied to
the $A_\varphi$ component of the gauge potential also leads to a regular form.

We will proceed in the following way.
At the $z$-axis we will expand the functions 
$H_i(r,\theta)$ and $\phi(r,\theta)$ in powers of $\sin\theta$. 
The coefficients will be functions of $r$ only. 
We insert the expansion of the functions $H_i(r,\theta)$ and $\phi(r,\theta)$ 
into the partial differential equations and expand the right hand sides of
Eqs.~(\ref{dgl_1})-(\ref{dgl_5}) in powers of $\sin\theta$. 
For a solution of the partial differential equations the
coefficients in the expansion of the partial differential equations 
have to vanish. 
Setting these coefficients equal to zero results 
in a set of relations for the coefficient functions of the expansion 
of the functions $H_i(r,\theta)$ and $\phi(r,\theta)$.
This does not determine the functions completely. 
However, 
it enables us to find the local form of
gauge transformations which lead to a gauge transformed potential 
which is well defined on the $z$-axis.

In a similar way we will expand the functions $H_i(r,\theta)$ and 
$\phi(r,\theta)$ at the origin in powers of $r\sin\theta$ and $r\cos\theta$. 
The coefficients in the expansion of the functions are constants then.
Relations between these constants will be obtained, again, 
by inserting the expansion of the functions into
the differential equations, 
expanding the expressions Eqs.~(\ref{dgl_1})-(\ref{dgl_5}) 
in powers of $r\sin\theta$ and $r\cos\theta$ 
and setting the coefficients in this expansion equal to zero. 
From this analysis the local form of a gauge transformation at the origin 
can be found, 
which again leads to a gauge transformed potential 
which is well defined at the origin.

Finally we will compare the gauge transformation at the origin  with 
the gauge transformations near the $z$-axis for small values of $r$.
As a result we will find, that the both local forms coincide in the region
where both expansions should hold.

Because our main interest is the behavior of the gauge potential at
the $z$-axis and at the origin, we will not exhibit the expansion of the 
dilaton function in the following. 

In this paper we will restrict to the case with winding number $n=2,3,4$.
The single steps in the calculations are elementary but tedious 
and will be omitted.
For winding number $n=3,4$ large expressions arise in the expansion at 
the origin. These expressions are exhibited in the Appendix. 

\subsection{Gauge Transformation at the $z$-axis}

We start by examining the behavior of the functions $H_i$ near the positive 
$z$-axis. We will expand these functions  $H_i$ in terms of $\sin\theta$.
From the equations of motion we will  
find relations between the coefficients which depend on $r$.
 
The expansions up to third order in $\theta$ was  
obtained before in ref.~\cite{MS-EYMD2} for the more general case
of Einstein-Yang-Mills-dilaton theory, i.~e. including the 
coupling of axially symmetric Yang-Mills and dilaton fields 
to gravity \cite{MS-EYMD1,MS-EYMD2,EYMDBH}. 
%Here we will need the expansions up to the fifth order.

\subsubsection*{n = 2}

The expansions in $\sin\theta$ to third order, which are consistent with the
boundary conditions Eqs.~(\ref{BC}), are given by
\begin{eqnarray}
H_1(r,\theta) &=& \tilde{H}_{11}(r) \sin\theta 
                  +\frac{\tilde{H}_{12}(r)}{2}\sin^2\theta 
                  +\frac{\tilde{H}_{13}(r)}{6}\sin^3\theta 
                  +O(\sin^4\theta)  \ ,
                 \nonumber\\
H_2(r,\theta) &=& f(r)  +\frac{\tilde{H}_{22}(r)}{2}\sin^2\theta 
                        +\frac{\tilde{H}_{23}(r)}{6}\sin^3\theta 
+O(\sin^4\theta) \ ,
                 \nonumber\\
H_3(r,\theta) &=& g(r) \sin\theta 
                  +\frac{\tilde{H}_{32}(r)}{2}\sin^2\theta 
                  +\frac{\tilde{H}_{33}(r)}{6}\sin^3\theta 
                  +O(\sin^4\theta)  \ ,
                 \nonumber\\
H_4(r,\theta) &=& f(r)  +\frac{\tilde{H}_{42}(r)}{2}\sin^2\theta 
                        +\frac{\tilde{H}_{43}(r)}{6}\sin^3\theta 
+O(\sin^4\theta) \ .
                 \nonumber\\
                 \label{expz_n2}
\end{eqnarray}
Inserting (\ref{expz_n2}) into the differential equations
(\ref{dgl_1})-(\ref{dgl_5}) and expanding in terms of $\sin\theta$
the following relations between the coefficient functions $\tilde{H}_{ij}(r)$
are found,
\begin{eqnarray}
\tilde{H}_{11} &=& -r\partial_r f\ ,
\nonumber                 \\
\tilde{H}_{22} &=& -r\partial_r (r\partial_r f)\label{rel_n2_22} \ ,
\nonumber                 \\
\tilde{H}_{23} &=& r\partial_r \tilde{H}_{12} \ ,
\nonumber                  \\
\tilde{H}_{32} &=& 0\ ,
\nonumber                 \\
\tilde{H}_{42} &=& \frac{1}{3}\left[2 f(f^2+ 3 g -1)
-\left[r\partial_r\right]^2  f\right]
\nonumber                  \ .
\end{eqnarray}
The functions $\tilde{H}_{33}(r)$ and $\tilde{H}_{43}(r)$
 may also 
be expressed in terms of $f(r)$, $g(r)$, $\tilde{H}_{12}(r)$,
$\tilde{H}_{13}(r)$ and other
functions. However these expressions look very complicated and their 
detailed structure is not needed in the following.

Expanding further $\sin\theta$ in terms of $\theta$ the 
functions $H_1$ and $H_2$ become
\begin{eqnarray}
H_1 &=& 
-r\partial_r f \theta 
+  \tilde{H}_{12} \frac{\theta^2}{2}
+ O(\theta^3) 
%\label{H1_n2_ex}\\
\nonumber\\
H_2 &=& 
 f
 -\left[r\partial_r\right]^2 f\frac{\theta^2}{2}
 +r\partial_r \tilde{H}_{12} \frac{\theta^3}{6} 
+ O(\theta^4)
\nonumber \\
   &=&
  \partial_\theta\left\{f\theta-\left[r\partial_r\right]^2 
  f\frac{\theta^3}{6}
  \right\}
 +r\partial_r \tilde{H}_{12} \frac{\theta^3}{6} 
+ O(\theta^4)
%\label{H2_n2_ex}\\
\nonumber
\end{eqnarray}
Comparing with Eqs.~(\ref{ET_H1}) and (\ref{ET_H2}) we see that a 
gauge transformation (\ref{U}) with 
\begin{eqnarray}
\Gamma = \Gamma^{(2)}_{(z)}(r,\theta) &\equiv& 
-\left\{ (f-1) \theta 
-\left[r\partial_r\right]^2 f \frac{\theta^3}{6}
\right\}
\label{GT_n2_z}\\
\nonumber
\end{eqnarray}
yields gauge transformed functions 
\begin{eqnarray}
\hat{H}_1 &=&  \tilde{H}_{12}  \frac{\theta^2}{2} 
                                                 + O(\theta^3)
             =  \frac{\tilde{H}_{12}}{2 r^2}\rho^2+ O(\rho^3)  \ ,          
%\label{hH1_n2_ex}\\
\nonumber\\
1-\hat{H}_2 &=&  -r\partial_r \tilde{H}_{12} \frac{\theta^3}{6} 
                            + O(\theta^4) 
             =  -\frac{r\partial_r \tilde{H}_{12}}{6r^3}\rho^3 
             + O(\rho^4)  \ ,                            
%\label{hH2_n2_ex}\\
\nonumber
\end{eqnarray}
which have the desired behavior on the positive $z$-axis, Eqs.~(\ref{reg-req}).

Performing the same gauge transformation to the functions $H_3$ and $H_4$
and calculating the gauge transformed functions $\hat{F}_3$ and $\hat{F}_4$,
we find  
\begin{eqnarray}
\hat{F}_3 &=& \sin\theta \hat{H}_3 +\cos\theta (1-\hat{H}_4)
=-\tilde{H}_{43}\frac{\theta^3}{6}  + O(\theta^4)=
 -\frac{\tilde{H}_{43}}{6r^3} \rho^3 + O(\rho^4) \ , 
%\label{hF3_n2_ex}\\
\nonumber\\
\hat{F}_4  &=& \cos\theta \hat{H}_3 -\sin\theta (1-\hat{H}_4)
= (f^2+2g-1)  \frac{\theta}{2} + O(\theta^3)
= \frac{f^2+2g-1}{2r} \rho + O(\rho^3)  \ ,  
%\label{hF4_n2_ex}\\
%
\nonumber
\end{eqnarray}
which also have the desired behavior on the positive $z$-axis, 
Eqs.~(\ref{reg-req}).

Turning to Cartesian coordinates we find for the  non-vanishing components
of the gauge potential, Eqs.~(\ref{A_cart}), to lowest order in the expansion

\begin{eqnarray}
\hat{A}_x &=&  
-\frac{x}{12r^4}\left[r\partial_r\tilde{H}_{12} 
- 3\tilde{H}_{12}\right]  \rho^2 \tau^2_\varphi 
-\frac{y}{6r^4}\tilde{H}_{43} \rho^2 \tau^2_\rho 
+\frac{y}{2r^2}\left[f^2+2g-1\right] \tau_3 \ ,
\nonumber\\
\hat{A}_y &=&  
-\frac{y}{12r^4}\left[r\partial_r\tilde{H}_{12} 
- 3\tilde{H}_{12}\right]  \rho^2 \tau^2_\varphi 
+\frac{x}{6r^4}\tilde{H}_{43} \rho^2 \tau^2_\rho 
-\frac{x}{2r^2}\left[f^2+2g-1\right] \tau_3 \ ,
\nonumber\\
\hat{A}_z &=&  
\frac{1}{4r^3} \tilde{H}_{12}\rho^2 \tau^2_\varphi  \ .
\nonumber
\end{eqnarray}

\subsubsection*{n = 3}

For winding number $n=3$ we consider the expansions in $\sin\theta$ to 
fourth order. They are given by
\begin{eqnarray}
H_1(r,\theta) &=& \tilde{H}_{11}(r) \sin\theta 
                  +\frac{\tilde{H}_{13}(r)}{6}\sin^3\theta 
                  +O(\sin^5\theta)  \ ,
                 \nonumber\\
H_2(r,\theta) &=& f(r)  +\frac{\tilde{H}_{22}(r)}{2}\sin^2\theta 
                        +\frac{\tilde{H}_{24}(r)}{24}\sin^4\theta 
+O(\sin^5\theta) \ ,
                 \nonumber\\
H_3(r,\theta) &=& g(r) \sin\theta 
                  +\frac{\tilde{H}_{33}(r)}{6}\sin^3\theta 
                  +O(\sin^5\theta)  \ ,
                 \nonumber\\
H_4(r,\theta) &=& f(r)  +\frac{\tilde{H}_{42}(r)}{2}\sin^2\theta 
                        +\frac{\tilde{H}_{44}(r)}{24}\sin^4\theta 
+O(\sin^5\theta) \ ,
                 \nonumber
%                 \label{expz_n3}
%
\end{eqnarray}
where we have omitted the vanishing terms.
The following relations between the coefficient functions 
are found from the expansion of 
the differential equations (\ref{dgl_1})-(\ref{dgl_5}),
\begin{eqnarray}
\tilde{H}_{11} &=& -r\partial_r f\ ,
\nonumber                 \\
\tilde{H}_{22} &=& -r\partial_r (r\partial_r f)\label{rel_n3_22} \ ,
\nonumber                 \\
\tilde{H}_{24} &=& 3 \tilde{H}_{22} + (r\partial_r \tilde{H}_{13}) \ ,
\nonumber                  \\
\tilde{H}_{42} &=& \frac{1}{3}\left[2 f(f^2+ 3 g -1)
-\left[r\partial_r\right]^2  f\right]
\nonumber                   \ .
\end{eqnarray}
The functions $\tilde{H}_{33}(r)$ and $\tilde{H}_{44}(r)$
 may also 
be expressed in terms of $f(r)$, $g(r)$, $\tilde{H}_{13}(r)$,
 and other
functions. 

Expanding $\sin\theta$ for small $\theta$ the functions $H_1$ and $H_2$
become
\begin{eqnarray}
H_1 &=& 
-r\partial_r f \theta 
+  (\tilde{H}_{13}+r\partial_r f ) \frac{\theta^3}{6}
+ O(\theta^5) \ ,
\nonumber\\
%\label{H1_n3_ex}\\
H_2 &=& 
 f
 -\left[r\partial_r\right]^2 f\frac{\theta^2}{2}
 +(r\partial_r\tilde{H}_{13}+\left[r\partial_r\right]^2 f)  
 \frac{\theta^4}{24} 
+ O(\theta^5) 
\nonumber \\
   &=&
  \partial_\theta\left\{f\theta-\left[r\partial_r\right]^2 f\
  \frac{\theta^3}{6}
  \right\}
 +(r\partial_r\tilde{H}_{13}+\left[r\partial_r\right]^2 f)
 \frac{\theta^4}{24} 
+ O(\theta^5)\ .
%\label{H2_n3_ex}\\
\nonumber
\end{eqnarray}
Comparing with Eqs.~(\ref{ET_H1}) and (\ref{ET_H2}) we see that for winding 
number $n=3$ a 
gauge transformation 
\begin{eqnarray}
\Gamma = \Gamma^{(3)}_{(z)}(r,\theta) &\equiv& 
-\left\{ (f-1) \theta 
-\left[r\partial_r\right]^2 f \frac{\theta^3}{6}
\right\}
%\label{GT_n3_z}\\
\nonumber
\end{eqnarray}
yields gauge transformed functions 
\begin{eqnarray}
\hat{H}_1 &=& 
( \tilde{H}_{13} + r\partial_r f-\left[r\partial_r\right]^3 f)
\frac{\theta^3}{6}  + O(\theta^5)
=                                                  
\frac{ \tilde{H}_{13} + r\partial_r f
-\left[r\partial_r\right]^3 f}{6r^3}\rho^3
                                                    + O(\rho^5)  \ ,  
\nonumber\\
%\label{hH1_n3_ex}\\
1-\hat{H}_2 &=&  -(r\partial_r \tilde{H}_{13}
+\left[r\partial_r\right]^2 f)
 \frac{\theta^4}{24} 
                            + O(\theta^5) 
=  -\frac{r\partial_r \tilde{H}_{13}
+\left[r\partial_r\right]^2 f}{24r^4}\rho^4
 + O(\rho^5)  \ , 
%\label{hH2_n3_ex}\\
\nonumber
\end{eqnarray}
which have the desired behavior on the positive $z$-axis,
 Eqs.~(\ref{reg-req}).

Performing the same gauge transformation to the functions $H_3$ and $H_4$
and calculating the gauge transformed functions $\hat{F}_3$ and $\hat{F}_4$ 
we find  
\begin{eqnarray}
\hat{F}_3 &=& \sin\theta \hat{H}_3 +\cos\theta (1-\hat{H}_4)
=
F_{33}  \sin^4\theta + O(\theta^5) 
=
\frac{F_{33}}{r^4}  \rho^4 + O(\rho^5) \ , 
\nonumber\\
%\label{hF3_n3_ex}\\
\hat{F}_4  &=& \cos\theta \hat{H}_3 -\sin\theta (1-\hat{H}_4)
= (f^2+2g-1)  \frac{\theta}{2} + O(\theta^3)
= \frac{f^2+2g-1}{2r} \rho + O(\rho^3)  \ ,  
%\label{hF4_n3_ex}\\
%
\nonumber
\end{eqnarray}
with
\begin{eqnarray}
F_{33} &=& 
\frac{1}{120}
\left(
20 g(2 f^3+f-\rdr^2 f)
+ 20 \tilde{H}_{33}f 
-5\tilde{H}_{44}
+4f(4f^4-4-5f\left[\rdr\right]^2 f)
\right) \ .
\nonumber
\end{eqnarray}
Again, these functions have the desired behavior on the positive $z$-axis, 
Eqs.~(\ref{reg-req}).

For winding number $n=3$ 
we find for the  non vanishing components
of the gauge potential in Cartesian coordinates 

\begin{eqnarray}
\hat{A}_x &=&  
-\frac{x}{48r^5}
\left[
r\partial_r \tilde{H}_{13}+\left[r\partial_r\right]^2 f
-4( \tilde{H}_{13} + r\partial_r f-\left[r\partial_r\right]^3 f)
\right]
\rho^3\tau^3_\varphi 
+\frac{3y}{2r^5} F_{33} \rho^3 \tau^3_\rho 
+\frac{3y}{4r^2}\left[f^2+2g-1\right] \tau_3 \ ,
\nonumber\\
\hat{A}_y &=&  
-\frac{y}{48r^5}
\left[
r\partial_r \tilde{H}_{13}+\left[r\partial_r\right]^2 f
-4( \tilde{H}_{13} + r\partial_r f-\left[r\partial_r\right]^3 f)
\right]
\rho^3\tau^3_\varphi 
-\frac{3x}{2r^5} F_{33} \rho^3 \tau^3_\rho 
-\frac{3x}{4r^2}\left[f^2+2g-1\right] \tau_3 \ ,
\nonumber\\
\hat{A}_z &=&  
\frac{1}{12r^4} \left[ \tilde{H}_{13} + r\partial_r f
-\left[r\partial_r\right]^3 f\right]
\rho^3 \tau^3_\varphi  \ ,
\nonumber
\end{eqnarray}
to lowest order in the expansion

\subsubsection*{n = 4}

Repeating the calculations of the last two subsections we
expand  the gauge field functions in $\sin\theta$ to fifth order,
\begin{eqnarray}
H_1(r,\theta) &=& \tilde{H}_{11}(r) \sin\theta 
                  +\frac{\tilde{H}_{13}(r)}{6}\sin^3\theta 
                  +\frac{\tilde{H}_{14}(r)}{24}\sin^4\theta 
                  +\frac{\tilde{H}_{15}(r)}{120}\sin^5\theta 
                  +O(\sin^6\theta)  \ ,
                 \nonumber\\
H_2(r,\theta) &=& f(r)  +\frac{\tilde{H}_{22}(r)}{2}\sin^2\theta 
                        +\frac{\tilde{H}_{24}(r)}{24}\sin^4\theta 
                        +\frac{\tilde{H}_{25}(r)}{120}\sin^5\theta 
+O(\sin^6\theta) \ ,
                 \nonumber\\
H_3(r,\theta) &=& g(r) \sin\theta 
                  +\frac{\tilde{H}_{33}(r)}{6}\sin^3\theta 
                  +\frac{\tilde{H}_{35}(r)}{120}\sin^5\theta 
                  +O(\sin^6\theta)  \ ,
                 \nonumber\\
H_4(r,\theta) &=& f(r)  +\frac{\tilde{H}_{42}(r)}{2}\sin^2\theta 
                        +\frac{\tilde{H}_{44}(r)}{24}\sin^4\theta 
                        +\frac{\tilde{H}_{45}(r)}{120}\sin^5\theta 
+O(\sin^6\theta) \ ,
                 \nonumber
%                 \label{expz_n4}
%
\end{eqnarray}
where again vanishing terms have been omitted.

From the expansion of the differential equations (\ref{dgl_1})-(\ref{dgl_5})
the following relations between the coefficient functions are found,
\begin{eqnarray}
\tilde{H}_{11} &=& -r\partial_r f\label{rel_n4_11} \ ,
\nonumber                \\
\tilde{H}_{13} &=&  -r\partial_r (f+ \tilde{H}_{22} )            
                 \nonumber\\
             &=& -\left\{r\partial_r f-\left[r\partial_r\right]^2 f\right\}\
%             \label{rel_n4_13}
             \ ,             
\nonumber                 \\
\tilde{H}_{22} &=& -r\partial_r (r\partial_r f)\label{rel_n4_22} \ ,
\nonumber                 \\
\tilde{H}_{24} &=& 4 \tilde{H}_{22}- r\partial_r (r\partial_r \tilde{H}_{22})
                 \nonumber\\
             &=& 
                 -\left\{4 \left[r\partial_r\right]^2 f 
                 -\left[r\partial_r\right]^4 f\right\}\ ,
%                 \label{rel_n4_24} 
\nonumber                  \\
\tilde{H}_{25}  &=& r\partial_r \tilde{H}_{14} \ ,
%                 \label{rel_n4_25} 
\nonumber                 \\
\tilde{H}_{42} &=& \frac{1}{3}\left[2 f(f^2+ 3 g -1)
-\left[r\partial_r\right]^2  f\right]
%                 \label{rel_n4_42} 
                 \ ,
\nonumber                 \\
\tilde{H}_{44} &=&  -\frac{1}{5}\left\{
                \left[r\partial_r\right]^2\tilde{H}_{22} 
                - 4 f^2 (4 f^3 + 10 f g + 5 \tilde{H}_{22} ) 
                - 4 f (5\tilde{H}_{33}  - 4 + 5 g)
                - 20 \tilde{H}_{22} g \right\}
                 \nonumber\\
             &=& \frac{1}{5}\left\{
                \left[r\partial_r\right]^4 f 
                +4 f^2 (4 f^3 + 10 f g - 5\left[r\partial_r\right]^2 f)
                +4 f (5\tilde{H}_{33}  - 4 + 5 g)
                -20  g\left[r\partial_r\right]^2 f \right\}\ .
%                 \label{rel_n4_44}
\nonumber                 
\end{eqnarray}
The functions $\tilde{H}_{15}(r)$, $\tilde{H}_{33}(r)$, $\tilde{H}_{35}(r)$,
$\tilde{H}_{45}(r)$ may also 
be expressed in terms of $f(r)$, $g(r)$, $\tilde{H}_{14}(r)$ and other
functions. 

Expanding $\sin\theta$ in terms of $\theta$ the functions $H_1$ and $H_2$
become
\begin{eqnarray}
H_1 &=& 
-r\partial_r f \theta 
+ \left[r\partial_r\right]^3 f \frac{\theta^3}{6}
+ \tilde{H}_{14}  \frac{\theta^4}{24} 
+ O(\theta^5) 
\nonumber \\
   &=&
- r\partial_r\left\{ f \theta -\left[r\partial_r\right]^2 f 
\frac{\theta^3}{6} \right\}
+ \tilde{H}_{14}  \frac{\theta^4}{24} 
+ O(\theta^5) \ ,
%\label{H1_n4_ex}
\nonumber\\
H_2 &=& 
 f
 -\left[r\partial_r\right]^2 f\frac{\theta^2}{2}
 +\left[r\partial_r\right]^4 f\frac{\theta^4}{24}
 +r\partial_r \tilde{H}_{14} \frac{\theta^5}{120} 
+ O(\theta^6)
\nonumber \\
   &=&
  \partial_\theta\left\{f\theta-\left[r\partial_r\right]^2 f\frac{\theta^3}{6}
  +\left[r\partial_r\right]^4 f\frac{\theta^5}{120}\right\}
 +r\partial_r \tilde{H}_{14} \frac{\theta^5}{120} 
+ O(\theta^6) \ .
%\label{H2_n4_ex}\\
\nonumber
\end{eqnarray}
As anticipated, we find  that for winding number $n=4$ a  
gauge transformation with 
\begin{eqnarray}
\Gamma= \Gamma^{(4)}_{(z)}(r,\theta) &\equiv& 
-\left\{ (f-1) \theta 
-\left[r\partial_r\right]^2 f \frac{\theta^3}{6}
+\left[r\partial_r\right]^4 f\frac{\theta^5}{120}\right\}
%\label{GT_n4_z}\\
\nonumber
\end{eqnarray}
yields gauge transformed functions 
\begin{eqnarray}
\hat{H}_1 &=&  \tilde{H}_{14}  \frac{\theta^4}{24} 
                                                 + O(\theta^5)
             =  \frac{\tilde{H}_{14}}{24 r^4}\rho^4+ O(\rho^5)  \ ,             
\nonumber
%\label{hH1_n4_ex}
\\
1-\hat{H}_2 &=&  -r\partial_r \tilde{H}_{14} \frac{\theta^5}{120} 
                            + O(\theta^6) 
             =  -\frac{r\partial_r \tilde{H}_{14}}{120r^5}\rho^5 + O(\rho^6)  
             \
             ,                            
%\label{hH2_n4_ex}\\
\nonumber
\end{eqnarray}
which have the desired behavior on the positive $z$-axis, Eqs.~(\ref{reg-req}).

Performing the same gauge transformation to the functions $H_3$ and $H_4$
and calculating the gauge transformed functions $\hat{F}_3$ and $\hat{F}_4$ 
we find  
\begin{eqnarray}
\hat{F}_3 &=& \sin\theta \hat{H}_3 +\cos\theta (1-\hat{H}_4)
=-\tilde{H}_{45}\frac{\theta^5}{120}  + O(\theta^6)=
 -\frac{\tilde{H}_{45}}{120r^5} \rho^5 + O(\rho^6) \ , 
%\label{hF3_n4_ex}
\nonumber
\\
\hat{F}_4  &=& \cos\theta \hat{H}_3 -\sin\theta (1-\hat{H}_4)
= (f^2+2g-1)  \frac{\theta}{2} + O(\theta^3)
= \frac{f^2+2g-1}{2r} \rho + O(\rho^3)  \ ,  
%\label{hF4_n4_ex}\\
%
\nonumber
\end{eqnarray}
which also have the desired behavior on the positive $z$-axis, 
Eqs.~(\ref{reg-req}).

Turning to Cartesian coordinates we find for the  non vanishing components
of the gauge potential to lowest order in the expansion

\begin{eqnarray}
\hat{A}_x &=&  
-\frac{x}{240r^6}\left[r\partial_r\tilde{H}_{14} - 5\tilde{H}_{14}\right] 
 \rho^4
\tau^4_\varphi 
-\frac{y}{60r^6}\tilde{H}_{45} \rho^4 \tau^4_\rho 
+\frac{y}{r^2}\left[f^2+2g-1\right] \tau_3 \ ,
\nonumber\\
\hat{A}_y &=&  
-\frac{y}{240r^6}\left[r\partial_r\tilde{H}_{14} - 5\tilde{H}_{14}\right]  
\rho^4
\tau^4_\varphi 
+\frac{x}{60r^6}\tilde{H}_{45} \rho^4 \tau^4_\rho 
-\frac{x}{r^2}\left[f^2+2g-1\right] \tau_3 \ ,
\nonumber\\
\hat{A}_z &=&  
\frac{1}{48r^5} \tilde{H}_{14}\rho^4 \tau^4_\varphi  \ .
\nonumber
\end{eqnarray}

The gauge transformation on the negative $z$-axis can be found in analogy 
to the analysis on the positive $z$-axis, taking into account the invariance 
of the differential equations (\ref{dgl_1})-(\ref{dgl_5}) 
under the transformation
\begin{equation}
\theta \rightarrow \pi - \theta \ , \
\dt  \rightarrow -\dt \ , \
H_1  \rightarrow -H_1 \ , \
H_3  \rightarrow -H_3 \ , \
H_2  \rightarrow  H_2 \ , \
H_4  \rightarrow  H_4 \ . \
\label{trans-z}
\end{equation}

\subsection{Gauge Transformation at the origin}\label{origin}
\noindent

The same strategy for finding a suitable gauge transformation
can be applied at the origin. It turns out, that for winding number $n$ 
an expansion to the order $n+3$ is required.

\subsubsection*{n = 2}

The analysis of the differential equations at the origin gives 
the following expressions for the expansion of the gauge field functions,
\begin{eqnarray}
H_{1}(r,\theta) 
& = &
r^2\sin\theta \cos\theta h_{12}
+
r^3\sin^2\theta \cos\theta h_{13} 
+
r^4\sin\theta \cos\theta\left(1-2 \sin^2\theta\right)h_{14}
%\nl
+
r^5\sin^2\theta \cos\theta\left(h_{15}+H_1^{(54)}\sin^2\theta\right) 
\nonumber\\
& &
\   +O(r^6) 
\nonumber\\
 & = &
 r\partial_r 
 \left\{ 
 \frac{1}{4} r^2 \sin(2\theta) h_{12}+\frac{1}{16} r^4 \sin(4\theta) h_{14}
 \right\}
\nonumber\\
& &
+
r^3\sin^2\theta \cos\theta h_{13} 
+
r^5\sin^2\theta \cos\theta\left(h_{15}+H_1^{(54)}\sin^2\theta\right)\   
+O(r^6) 
\ , \nonumber\\ 
& &
\nonumber\\ 
H_{2}(r,\theta)
& = &
1
-
\frac{1}{2}r^2\left(1-2\sin^2\theta\right) h_{12}
+
r^3\sin^3\theta h_{13} 
-
\frac{1}{4}r^4 h_{14}\left(1-8\sin^2\theta+8 \sin^4\theta \right)
%\nl
\nonumber\\
& &
+
\frac{1}{3}r^5\sin^3\theta \left(5 h_{15}+3 H_1^{(54)}\sin^2\theta\right)
\   +O(r^6) 
\nonumber\\ 
 & = &
 -\partial_\theta
 \left\{ 
 \frac{1}{4} r^2 \sin(2\theta) h_{12}+\frac{1}{16} r^4 \sin(4\theta) h_{14}
 \right\}
\nonumber\\
& &
+
1
+
r^3\sin^3\theta h_{13} 
+
\frac{1}{3}r^5\sin^3\theta \left(5 h_{15}+3 H_1^{(54)}\sin^2\theta\right)
\  +O(r^6) 
\ , \nonumber\\ 
& &
\nonumber\\
H_{3}(r,\theta) & = & 
r^2\sin\theta \cos\theta h_{32}
+
r^4\sin\theta \cos\theta \left(h_{34c} +h_{34s}\sin^2\theta\right) 
%\nl
+
r^5\sin^4\cos\theta H_3^{(54)}\  +O(r^6) 
\ , \nonumber\\
H_{4}(r,\theta) & = & 
1-\frac{1}{2}r^2\left(h_{12}-2h_{32}\sin^2\theta\right)
+
r^4 \left(-\frac{1}{4}h_{14}  +H_4^{(42)}\sin^2\theta 
+h_{34s}\sin^4\theta\right)
%\nl
-
r^5 \sin^3\theta \cos^2\theta H_3^{(54)}\  +O(r^6) 
\ , \nonumber\\
 \label{expo_n2}
\end{eqnarray}
with
\begin{eqnarray}
H_3^{(54)}
& = &
\frac{1}{2}e^{2 \phi_0}  
        h_{13} 
        \left(h_{12}-2h_{32}\right)^{2}
  +    h_{13s} \phi^{(2)} 
  +
  \frac{1}{12}h_{15}
\ , \nonumber\\
H_4^{(42)}
& = &
\frac{1}{4}\left(
         h_{12}^{2}
        -2 h_{12} h_{32}
        +2 h_{14} 
        +4 h_{34c}
        \right)
\ , \nonumber\\
H_1^{(54)}
& = &
-\frac{1}{6}e^{2 \phi_0} h_{13} 
        \left(h_{12}-2 h_{32}\right)^{2}
  +
  \frac{1}{3}  h_{13} \phi^{(2)} 
  -
 \frac{1}{12}\left(
        23 h_{15}
        +4  h_{13}\left(h_{12} -2 h_{32}\right)
        \right)
\ , \nonumber\\
 \phi^{(2)} 
& = &
\frac{2}{   \left(h_{12}-2 h_{32}\right)}
 \left( \frac{1}{32}  \left(-9 h_{12}^{2}+16 h_{12} h_{32}-6 h_{14}-40
 h_{34c}\right)
-h_{34s}\right)
\nonumber
% \label{expo2}
 \end{eqnarray}
and where $\phi_0$ and $h_{ij}$  are constants.

It can be seen easily, that a gauge transformation with 
$$
%\begin{equation}
\Gamma = \Gamma^{(2)}_{(o)}(r,\theta) =
 \left\{ 
 \frac{1}{4} r^2 \sin(2\theta) h_{12}+\frac{1}{16} r^4 \sin(4\theta) h_{14}
 \right\}\  +O(r^6) 
%\label{gam_o}
%\end{equation}
$$
removes the low order terms in $H_1$ and $1-H_2$.
The gauge transformed functions become
\begin{eqnarray}
\hat{H}_1 &=&
r^3\sin^2\theta \cos\theta h_{13} 
+
r^5\sin^2\theta \cos\theta\left(h_{15}+H_1^{(54)}\sin^2\theta\right)
 \  +O(r^6) 
\nonumber \\
&=&
z \rho^2\left[
h_{13} + z^2 h_{15}+\rho^2 (H_1^{(54)}+ h_{15})
\right]\  +O(r^6) 
\ ,  
%\label{hH1_o}
\nonumber\\
1-\hat{H}_2 &=&
-r^3\sin^3\theta h_{13} 
-
\frac{1}{3}r^5\sin^3\theta \left(5 h_{15}+3 H_1^{(54)}\sin^2\theta\right)
 \    +O(r^6) 
\nonumber \\
&=&
-\rho^3\left[h_{13}+\frac{1}{3}(5 z^2 h_{15}+\rho^2( 3 H_1^{(54)}+5 h_{15} ))
\right]\  +O(r^6) 
\ . 
% \label{hH2_o}
\nonumber 
\end{eqnarray}
 
Performing the same gauge transformation to the functions $H_3$ and $H_4$
we find for the functions $\hat{F}_3$ and $\hat{F}_4$
\begin{eqnarray}
\hat{F}_3 &=& 
\frac{1}{12}z r \rho^3 F_{33o}\ + O(r^6) \ ,
%\label{F3_o}
\nonumber \\
\hat{F}_4 &=& 
-\frac{1}{2}r \rho (h_{12}-2h_{32}) \ + O(r^4) \ ,
%\label{F4_o}
\nonumber 
\end{eqnarray}
with
\begin{eqnarray}
F_{33o}
 &=&\frac{1}{12}  \left[
h_{15}+12   h_{13} \phi^{(2)} 
+ 6  e^{2 \phi_0}  h_{13}(h_{12}-2h_{32})^2\right]\ .
\nonumber
\end{eqnarray}
Thus the gauge transformed functions fulfill the requirement,
Eqs.~(\ref{reg-req}) up to order $r^5$.

Near the origin the components of the transformed gauge potential 
in Cartesian coordinates become
\begin{eqnarray}
\hat{A}_x &=& 
-\frac{1}{3} z x h_{15} \rho^2 \tau^2_\varphi + z y F_{33o}\rho^2 \tau^2_\rho
-\frac{1}{2} y (h_{12}-2 h_{32}) \tau_3 \ ,
\nonumber\\
\hat{A}_y &=& 
-\frac{1}{3} z y h_{15} \rho^2 \tau^2_\varphi - z x F_{33o}\rho^2 \tau^2_\rho
+\frac{1}{2} x (h_{12}-2 h_{32}) \tau_3 \ ,
\nonumber\\
\hat{A}_z &=&
\frac{1}{2} h_{13} \rho^2 \tau^2_\varphi \ , 
\nonumber
\end{eqnarray}
to leading order in the expansion Eq.~(\ref{expo_n2}).

\subsubsection*{n = 3 and n = 4 }

The same analysis can be performed for higher 
winding numbers. In this section we give the results 
for winding number $n=3$ and $n=4$.
For winding number  $n=3$ the expansion to sixth order is found to be
\begin{eqnarray}
H_{1}(r,\theta) 
& = &
r^2\sin\theta \cos\theta h_{12}
+
r^4\sin\theta \cos\theta
\left(h_{14s} \sin^2\theta+h_{14c} \cos^2\theta\right)
\nonumber\\
& &
+
r^6\sin\theta \cos\theta
\left(a_{11} +a_{13} \sin^2\theta +H_1^{(65)} \sin^4\theta\right) \ ,
\nonumber\\
H_{2}(r,\theta) 
& = &
1
-
\frac{1}{2}r^2(1-2\sin^2\theta) h_{12}
-
\frac{1}{4}r^4 
\left(h_{14c}-8h_{14c}\sin^2\theta+4(h_{14c}-h_{14s})\sin^4\theta\right)
\nonumber\\
& &
-
\frac{1}{6}r^6
\left(a_{11} -18 a_{11} \sin^2\theta 
                  -9 a_{13} \sin^4\theta -6 H_1^{(65)} \sin^6\theta\right) \ ,
\nonumber\\
H_{3}(r,\theta) 
& = &
r^2\sin\theta \cos\theta h_{32}
+
r^4\cos\theta 
\left(H_3^{(41)}\sin\theta +H_3^{(43)}\sin^3\theta\right) 
\nonumber\\
& &
+
r^6\cos\theta 
\left(H_3^{(61)}\sin\theta +H_3^{(63)}\sin^3\theta
                +H_3^{(65)}\sin^5\theta\right)  \ ,
\nonumber\\
H_{4}(r,\theta)
& = &
1-\frac{1}{2}r^2
\left(h_{12}-2h_{32}\sin^2\theta\right)
+
r^4 
\left(-\frac{1}{4}h_{14c} +H_4^{(42)}\sin^2\theta 
+H_3^{(43)}\sin^4\theta\right)
\nonumber\\
& &
+
r^6 
\left(-\frac{1}{6}a_{11} +H_4^{(62)}\sin^2\theta 
     +H_4^{(64)}\sin^4\theta +H_4^{(66)}\sin^6\theta\right) \ ,
\nonumber\\
 \label{expo_n3}
 \end{eqnarray}
where $h_{ij}$, $a_{ij}$, 
and  $H_i^{(jk)}$
are constants. The details of the expressions  $H_i^{(jk)}$ 
are given in the Appendix.

For $n=3$ the gauge transformation function becomes
$$
%\begin{equation}
\Gamma = \Gamma^{(3)}_{(o)}(r,\theta) = 
 \left\{ 
  \frac{1}{4} r^2 \sin(2\theta) h_{12}
 +\frac{1}{16} r^4 \sin(4\theta) h_{14c}
 +\frac{1}{36} r^6 \sin(6\theta) a_{11} 
 \right\}\  +O(r^7) \ . 
%\label{gam_o_n3}
%\end{equation}
$$

For the gauge transformed functions we find
\begin{eqnarray}
\hat{H}_1 &=&
z\rho^3 (h_{14s}+h_{14c}) 
+
\frac{1}{3}z\rho^3 
\left(
          3(a_{13}+H_1^{(65)})\rho^2
         +(16a_{11}+3a_{13})z^2 
\right)
 \  +O(r^7) 
\ , 
% \label{hH1_n3_o}
\nonumber\\
1-\hat{H}_2 &=&
-\rho^4 (h_{14s}+h_{14c}) 
-\rho^4 
\frac{1}{6}\left((16a_{11}+3(3a_{13}+2 H_1^{(65)}))\rho^2
                 +3(16a_{11}+3a_{13})z^2\right)
  +O(r^7) 
\ ,  
%\label{hH2_o_n3}
\nonumber\\
\hat{F}_3 &=& 
z r \rho^4 F_{34o}\ + O(r^7) \ ,
%\label{F3_o_n3}
\nonumber\\
\hat{F}_4 &=& 
-\frac{1}{2}r \rho (h_{12}-2h_{32}) \ + O(r^4) \ ,
%\label{F4_o_n3}
\nonumber
\end{eqnarray}
where $F_{34o}$ is a constant given by
\begin{eqnarray}
F_{34o}
 &=&\frac{1}{90}  \left[
16 a_{11}+3 a_{13}+48   (h_{14s}+h_{14c}) \phi^{(2)} 
+ 54  e^{2 \phi_0}(h_{14s}+h_{14c}) (h_{12}-2h_{32})^2\right]\ .
\nonumber
\end{eqnarray}

Near the origin the components of the
transformed gauge potential in Cartesian coordinates become
\begin{eqnarray}
\hat{A}_x &=& 
-\frac{1}{12} z x (16 a_{11}+3 a_{13}) \rho^3 \tau^3_\varphi 
+\frac{3}{2} z y F_{34o}\rho^3 \tau^3_\rho
-\frac{3}{4} y (h_{12}-2 h_{32}) \tau_3 \ ,
\nonumber\\
\hat{A}_y &=& 
-\frac{1}{12} z y (16 a_{11}+3 a_{13}) \rho^3 \tau^3_\varphi 
-\frac{3}{2} z x F_{34o}\rho^3 \tau^3_\rho
+\frac{3}{4} x (h_{12}-2 h_{32}) \tau_3 \ ,
\nonumber\\
\hat{A}_z &=&
\frac{1}{2}(h_{14s}+h_{14c}) \rho^3 \tau^3_\varphi \ , 
\nonumber
\end{eqnarray}
to leading order in the expansion Eq.~(\ref{expo_n3}).

For winding number  $n=4$ the expansion to seventh order is found to be
\begin{eqnarray}
H_{1}(r,\theta) 
& = &
r^2\sin\theta \cos\theta h_{12}
-
r^4\sin\theta \cos\theta h_{14}(1-2 \sin^2\theta)
\nonumber\\
& &
+
r^5\sin^4\theta\cos\theta h_{15}
+
\frac{1}{3}a_{11} r^6\sin\theta \cos\theta
(3-16\sin^2\theta+16\sin^4\theta)
+
r^7 H_1^{(7)}\sin^4\theta \cos\theta \ ,
\nonumber\\
H_{2}(r,\theta) 
& = &
1
-
\frac{1}{2}r^2(1-2\sin^2\theta) h_{12}
+
\frac{1}{4}r^4 h_{14} (1-8\sin^2\theta+8\sin^4\theta )
\nonumber\\
& &
+
r^5\sin^5\theta h_{15}
-
\frac{1}{6}a_{11}r^6
\left(1-18\sin^2\theta +48\sin^4\theta-32\sin^6\theta\right)
+
r^7 \sin^5\theta H_2^{(7)}\ ,
\nonumber\\
H_{3}(r,\theta) 
& = &
r^2\sin\theta \cos\theta h_{32}
+
r^4\cos\theta 
\left(H_3^{(41)}\sin\theta +H_3^{(43)}\sin^3\theta\right) 
\nonumber\\
& &
+
r^6\cos\theta 
\left(H_3^{(61)}\sin\theta +H_3^{(63)}\sin^3\theta
                +H_3^{(65)}\sin^5\theta\right) 
+
r^7 H_3^{(7)}\sin^6\theta \cos\theta\ ,
\nonumber\\
H_{4}(r,\theta) 
& = &
1-\frac{1}{2}r^2(h_{12}-2h_{32}\sin^2\theta)
+
r^4 
\left(\frac{1}{4}h_{14} +H_4^{(42)}\sin^2\theta +H_3^{(43)}\sin^4\theta\right)
\nonumber\\
& &
+
r^6 
\left(-\frac{1}{6}a_{11} +H_4^{(62)}\sin^2\theta 
     +H_4^{(64)}\sin^4\theta +H_3^{(65)}\sin^6\theta\right)
-
r^7 H_3^{(7)}\sin^5\theta \cos^2\theta\ ,
\nonumber\\
 \label{expo_n4}
 \end{eqnarray}
with
\begin{eqnarray}
H_1^{(7)}
&=&
-e^{2  \phi_0} h_{15} \cos^2\theta 
  (h_{12}-2 h_{32})^{2}
  +    h_{15}  \phi^{(2)}  \cos^2\theta    
\nonumber\\
& &
  +
    \frac{1}{8}\left(
    23  d_{12} \sin^{2}\theta  
      -15  d_{12}
      -12  h_{15}(h_{12}-2 h_{32}) \cos^2\theta 
    \right)\ ,
\nonumber\\
H_2^{(7)}
&=&
-\frac{1}{5}e^{2  \phi_0} h_{15}(7-5\sin^2\theta) 
  (h_{12}-2 h_{32})^{2}
  +\frac{1}{5}    h_{15}  \phi^{(2)}  (7-5\sin^2\theta)   
\nonumber\\
& &
  -
    \frac{3}{10}
        h_{15}(h_{12}-2h_{32}) (7-5\sin^2\theta)  
    -\frac{1}{8}d_{12}(21-23\sin^2\theta)\ ,
\nonumber\\
H_3^{(7)}
&=&
\frac{13}{20}  e^{2  \phi_0}  
          h_{15}     
        (h_{12}-2  h_{32})^{2}
%\nonumber\\
%& &
  +
  \frac{7}{20}       h_{15}  \phi^{(2)}  
  -
  \frac{1}{160}  
        \left(5  d_{12}+
          4   h_{15}(h_{12}-2  h_{32})
        \right)\ ,
\nonumber
\end{eqnarray}
where $h_{ij}$, $a_{11}$, $d_{12}$, $\phi^{(2)}$,
$\phi_0$
and the  $H_i^{(jk)}$
are constants. The details of the expressions  $H_i^{(jk)}$ 
are given in the Appendix.

For $n=4$ the gauge transformation function becomes
\begin{equation}
\Gamma= \Gamma^{(4)}_{(o)}(r,\theta) = 
 \left\{ 
  \frac{1}{4} r^2 \sin(2\theta) h_{12}
 -\frac{1}{16} r^4 \sin(4\theta) h_{14}
 +\frac{1}{36} r^6 \sin(6\theta) a_{11} 
 \right\}\  +O(r^7) \ . 
\label{gam_o_n4}
\end{equation}

Near the origin the transformed gauge potential in Cartesian coordinates 
becomes
\begin{eqnarray}
\hat{A}_x &=& 
 z x F_{15o} \rho^4 \tau^4_\varphi 
+ z y F_{35o}\rho^4 \tau^4_\rho
- y (h_{12}-2 h_{32}) \tau_3 \ ,
\nonumber\\
\hat{A}_y &=& 
 z y F_{15o} \rho^4 \tau^4_\varphi 
- z x F_{35o}\rho^4 \tau^4_\rho
+ x (h_{12}-2 h_{32}) \tau_3 \ ,
\nonumber\\
\hat{A}_z &=&
\frac{1}{2}h_{15} \rho^4 \tau^4_\varphi \ , 
\nonumber
\end{eqnarray}
with
\begin{eqnarray}
 F_{15o}
 &=& 
\frac{1}{40}
\left(
15 d_{12} +4 h_{15}
 \left[3(h_{12}-2h_{32})-2   \phi^{(2)}
          +2 e^{2  \phi_0}(h_{12}-2 h_{32})^2
 \right]         
\right) \ , 
\nonumber \\
 F_{35o}
 &=& 
-\frac{1}{80}
\left(
5 d_{12} +4 h_{15}
 \left[h_{12}-2h_{32}-14   \phi^{(2)}
 -26  e^{2  \phi_0}(h_{12}-2 h_{32})^2
  \right] 
\right) \ , 
 \nonumber
\end{eqnarray}
to leading order in the expansion Eq.~(\ref{expo_n4}).

As for winding number $n=2$ the gauge transformed functions 
for $n=3$ and $n=4$ fulfill the requirement,
Eqs.~(\ref{reg-req}) up to order $r^5$ and  $r^7$, respectively.

\subsection{Gauge Transformation on the overlap}
\noindent

Next we compare the gauge transformations $\Gamma^{(n)}_{(z)}$ on the 
positive $z$-axis and $\Gamma^{(n)}_{(o)}$ at the origin.
By comparison of the expansions of the functions 
$H_i(r,\theta)$ near the $z$-axis and near the origin we can determine
the expansion of the function $f(r)$ at the origin and calculate the
gauge transformation function $\Gamma_{(z)}(r,\theta)$ for small 
values of $r$, $\theta$.
For winding number $n=4$ for example, we find
\begin{eqnarray}
\Gamma^{(4)}_{(z)}(r,\theta)
&=&
(-\frac{1}{2} r^2 h_{12} +\frac{1}{4} r^4 h_{14} -\frac{1}{6}r^6 a_{11} )
            \theta \nonumber \\
& &
-(-2 r^2 h_{12} + 4 r^4 h_{14} - 6  r^6 a_{11} )    
           \frac{1}{6} \theta^3 \nonumber \\
& &
+(-8 r^2 h_{12} + 64 r^4 h_{14} - 216  r^6 a_{11} ) 
           \frac{1}{120} \theta^5 \nonumber \\
&=&
-\frac{1}{4}r^2  h_{12}
( 2\theta - \frac{1}{6} (2\theta)^3 + \frac{1}{120} (2\theta)^5)
 \nonumber \\          
& &
+\frac{1}{16}r^4 h_{14}
( 4\theta - \frac{1}{6} (4\theta)^3+ \frac{1}{120} (4\theta)^5)
 \nonumber \\          
& &
-\frac{1}{36}r^6  a_{11}
( 6\theta - \frac{1}{6} (6\theta)^3+ \frac{1}{120} (6\theta)^5) \ .
 \nonumber         
\end{eqnarray}
The same expression can be obtained by expanding 
$\Gamma^{(4)}_{(o)}(r,\theta)$,
Eq.~(\ref{gam_o_n4}),
for small values of $\theta$.
Thus both gauge transformations coincide on the neighborhood where
the expansion for small $\theta$ and small $r$ are comparable.
The same holds for winding number $n=2,3$.

Similarly, the gauge transformation function on the negative $z$-axis can 
be compared with $\Gamma^{(n)}_{(o)}(r,\theta)$. 
Again the both gauge transformations
coincide where the expansions are comparable.

\section{Discussion and Conclusions}\label{conclusion}

In this paper we have examined the question as to how  static 
axially symmetric solutions of the Yang-Mills-dilaton theory, 
given within a singular Ansatz of the gauge potential, 
can be gauge transformed into regular form.
We have shown that the field strength tensor can be calculated 
straightforwardly from the singular Ansatz, if we impose the
condition of finite energy density on the gauge field functions.
Although the field strength tensor
itself is not well defined on the singular axis and the origin, 
the Lagrange density is well defined in terms of the 
gauge field functions. Consequently the equations of motion reduce
to a boundary value problem for the gauge field functions and the 
dilaton function.
From the analysis of this boundary value problem we found that the
singular parts of the 
$A_r$ and $A_\theta$ components of the gauge potential behave 
like a pure gauge along the singular axis and at the origin.
With suitable local gauge transformations the non-regular parts
have been gauged away. 
Applying the same gauge transformation to the $A_\varphi$ component of
the singular gauge potential leads to a regular form, too.
Thus in total the gauge transformed gauge potential is regular.

The gauge transformation itself is not well defined. 
However, it does not introduce
new terms in the field strength tensor. 
It can be checked, that 
along the singular axis and at the origin the
field strength tensor $\hat{{\cal F}}_{\mu\nu}[\hat{A}]$,
calculated from the regular gauge transformed potential $\hat{A}$,
coincides with 
the gauge rotated field strength tensor $U{\cal F}_{\mu\nu}[A]U^{\dagger}$
in terms of the singular gauge
potential $A$,
provided the gauge field functions are used in their local
expansions, i.~e. for solutions of the boundary value problem.

The gauge transformation functions $\Gamma^{(n)}_{(z)}$,  
$\Gamma^{(n)}_{(o)}$ are determined only in the 
vicinity of the $z$-axis or the origin, respectively, 
and have to be considered as local gauge transformations. 
Away form the $z$-axis or the origin the gauge transformation 
matrices $U_{(z)}$, $U_{(o)}$ might approach unity sufficiently fast. 
Then the gauge transformed functions $\hat{H}_i$ 
behave like the original functions $H_i$ except in a region near the
$z$-axis or the origin.

The gauge transformation derives from the behavior of the 
gauge field functions on the singular axis and at the origin which in turn
depend on the theory under consideration. 
Thus, the gauge transformations given in this paper for
Yang-Mills-dilaton theory do not lead in general
to regular gauge potentials in different theories involving 
the static axially symmetric Ansatz, Eq.~(\ref{An}). 
However, we have outlined a possible way how one can proceed
to find suitable gauge transformations in theories, 
where multi-monopoles \cite{KKT,KOT2} or multi-sphalerons
\cite{MS_WS,MS-EYMD1,MS-EYMD2,BraVar,EYMDBH} have been constructed
within the Ansatz Eq.~(\ref{An}).
For example,  
for the multi-sphalerons in Einstein-Yang-Mills(-dilaton)
theory the expansion on the positive $z$-axis up to third
order in $\theta$ is given in ref.~\cite{MS-EYMD2}.
From their result it can be seen easily, that for winding number $n=2$ 
the gauge transformation leading to a regular gauge potential along the 
positive $z$-axis coincides with the corresponding gauge transformation
given in this paper, Eq.~(\ref{GT_n2_z}).
We conjecture, that the same result will 
hold for the static axially symmetric black hole solutions in 
Einstein-Yang-Mills(-dilaton) theory constructed in ref.~\cite{EYMDBH}.
More work has to be done for the 
sphaleron and black hole solutions carrying 
larger winding numbers \cite{future}.

Finally let us compare with the spherically symmetric case. The Ansatz
for the gauge potential
$$
 A_i(\vec{r}) = \frac{a(r)-1}{2r} \epsilon_{ij\alpha} 
                                              \frac{r^j}{r}\tau_\alpha
$$
is not well defined at the origin, $r=0$, even when the boundary condition
$a(0)=1$ is imposed. Regularity requires $a(r)=1-b r^2 +O(r^3)$, where $b$
is a constant. This condition is not guaranteed a priori.
Instead, for a given model it has to be checked whether the 
regularity condition follows from the analysis of the solution at the 
origin, i.~e. at the singular point. The similarity to the axially 
symmetric case is striking. In that case we have also performed the 
analysis at the singularities, the $z$-axis and the origin, and found 
that the gauge potential is regular, if we apply  suitable gauge
transformations.

\bigskip

{\bf Acknowledgments}
The author thanks Y. Brihaye, P. Kosinski and J. Kunz for 
helpfull discussions.
This work was carried out under 
Basic Science Research project SC/97/636 of
FORBAIRT.

\newpage

\section*{Appendix}\label{H_ijk}

In this section the expressions $H_i^{(jk)}$ used in section \ref{origin}
for winding number $n=3$ and $n=4$ are given.

\subsection*{n = 3}

\begin{eqnarray}
H_1^{(65)}& = &
\frac{1}{80}\left(9\left(h_{14c}+h_{14s}\right)\left(
 4  \phi^{(2)}-5 (h_{12} -2h_{32})
- 3  e^{2 \phi_0} \left(h_{12}-2h_{32}\right)^2\right)
-2\left(144 a_{11}+67a_{13}\right)\right)
\nonumber 
\end{eqnarray}
\begin{eqnarray}
H_3^{(43)}& = &
 2\phi^{(2)}  
  \left(h_{12}
    -2  h_{32}
  \right)
  +
  \frac{1}{8}   
        \left(9  h_{12}^{2}
          -16  h_{12}  h_{32}
          +6  h_{14c}
          +40  h_{34}
        \right)
\nonumber 
\end{eqnarray}
\begin{eqnarray}
H_3^{(41)}& = &
 -2 \phi^{(2)}  
  \left(h_{12}
    -2  h_{32}
  \right)
  -
  \frac{1}{8}   
        \left(
          9  h_{12}^{2}
          -16  h_{12}  h_{32}
          +6  h_{14c}
          +32  h_{34}
        \right)
\nonumber 
\end{eqnarray}
\begin{eqnarray}
H_3^{(65)}& = &
 \frac{3}{2}  e^{2 \phi_0}  
        \phi^{(2)}  
        \left(h_{12}-2 h_{32}\right)^{3}
%        \nonumber \\ & &
  -12  e^{2 \phi_0} 
  \left(h_{12}-2 h_{32}\right)^2\left(h_{14c}+h_{14s}\right)
%        \nonumber \\ & &
  +
  \frac{4}{3}      (\phi^{(2)})^{2}  
        \left(h_{12}-2  h_{32}\right)
        \nonumber \\ & &
 - \frac{1}{3}      
    \left(
     \left( 5  h_{12}\left(h_{12}-2 h_{32}\right)  
      +40  h_{14c}  \
      +32  h_{34}  
      +32  h_{14s}   
       \right)
  \phi^{(2)}
   +28 \left(h_{12}-2 h_{32}\right)\phi_{41}
    \right)
        \nonumber \\ & &
  +
  \frac{1}{24}   
    \left(-
      148  a_{11}
      -16  a_{13}
      +504  c_{32}
      -49  h_{12}^{3}
      +99  h_{12}^{2}  h_{32}
%      \right.
%        \nonumber \\ & &
%        \left.
      -111  h_{12}  h_{14c}
      -192  h_{12}  h_{34}
      +114  h_{14c}  h_{32}
    \right)
\nonumber 
\end{eqnarray}
\begin{eqnarray}
H_3^{(63)}& = &
 -3  e^{2 \phi_0}     
  \phi^{(2)}  
  \left(h_{12}-2 h_{32}\right)^{3}
%        \nonumber \\ & &
  +
  \frac{84}{5}  e^{2 \phi_0}     
    \left(h_{12}-2 h_{32}\right)^{2}\left(h_{14c}+h_{14s}\right)
%        \nonumber \\ & &
  -
  \frac{8}{3}(\phi^{(2)})^{2}  
        \left(h_{12} -2  h_{32}\right)
        \nonumber \\ & &
  +
  \frac{1}{15}      
    \left(
    \left(
    35  h_{12}\left(h_{12}-2h_{32}\right)  
      +274  h_{14c}  
      +224  h_{14s}  
      +200  h_{34} 
    \right)
       \phi^{(2)}
       +220
    \left(h_{12}-2h_{32}\right)\phi_{41}
    \right)
        \nonumber \\ & &
  +
   \frac{1}{360}
    \left(3152  a_{11}
      +336  a_{13}
      -10080  c_{32}
      +1095  h_{12}^{3}
      -2205  h_{12}^{2}  h_{32}
%      \right.
%        \nonumber \\ & &
%        \left.
      +2430  h_{12}  h_{14c}
      +4320  h_{12}  h_{34}
      -2430  h_{14c}  h_{32}
    \right)
\nonumber 
\end{eqnarray}
\begin{eqnarray}
H_3^{(61)}& = &
 \frac{3}{2}  e^{2 \phi_0}     
        \phi^{(2)}  
        \left(h_{12}-2 h_{32}\right)^{3}
%        \nonumber \\ & &
  -
  \frac{24}{5}  e^{2 \phi_0}     
    \left(
      h_{12}  -2 h_{32}\right)^{2}\left(h_{14c}+h_{14s}\right)
%        \nonumber \\ & &
  +
  \frac{4}{3}  (\phi^{(2)})^{2}  
        \left(h_{12}-2  h_{32}\right)
        \nonumber \\ & &
  -
  \frac{2}{15}         
    \left(
    \left(
      5  h_{12}\left( h_{12}-2  h_{32}\right) 
      +37  h_{14c}  \
      +32  h_{14s}  \
      +20  h_{34}  
    \right)
      \phi^{(2)}
       +40
    \left(h_{12}-2  h_{32}  \right)
      \phi_{41}
    \right)
        \nonumber \\ & &
  +
  \frac{1}{360}   
    \left(
      -932  a_{11}
      -96  a_{13}
      +2880  c_{32}
      -360  h_{12}^{3}
      +720  h_{12}^{2}  h_{32}
%      \right.
%        \nonumber \\ & &
%        \left.
      -765  h_{12}  h_{14c}
      -1440  h_{12}  h_{34}
      +720  h_{14c}  h_{32}
    \right)
\nonumber 
\end{eqnarray}
\begin{eqnarray}
H_4^{(42)}& = &
 -2         \phi^{(2)}  
  \left(h_{12}
    -2  h_{32}
  \right)
  -
  \frac{1}{8}   
        \left(
          7  h_{12}^{2}
          -12  h_{12}  h_{32}
          +2  h_{14c}
          +32  h_{34}
        \right)
\nonumber 
\end{eqnarray}
\begin{eqnarray}
H_4^{(66)}& = &
 \frac{3}{2}  e^{2 \phi_0}  \phi^{(2)}  
        \left(h_{12}-2  h_{32}\right)^{3}
%        \nonumber \\ & &
  -12  e^{2 \phi_0}  
  \left(h_{12}-2  h_{32}\right)^2\left(h_{14c}+h_{14s}\right)
%        \nonumber \\ & &
  +
  \frac{4}{3}  (\phi^{(2)})^{2}  
        \left(h_{12}-2  h_{32}\right)
        \nonumber \\ & &
  -
  \frac{1}{3}   
    \left(
    \left(
      5  h_{12}\left(h_{12}-2  h_{32}\right)  
      +40  h_{14c}  
      +32  h_{14s}  
      +32  h_{34} 
    \right)
      \phi^{(2)}
      +28
    \left(h_{12}-2  h_{32}\right)
      \phi_{41}
    \right)
        \nonumber \\ & &
  +
  \frac{1}{24}\left(
    -148  a_{11}
    -16  a_{13}
    +504  c_{32}
    -49  h_{12}^{3}
    +99  h_{12}^{2}  h_{32}
%    \right.
%        \nonumber \\ & &
%        \left.
    -111  h_{12}  h_{14c}
    -192  h_{12}  h_{34}
    +114  h_{14c}  h_{32}
  \right)
\nonumber 
\end{eqnarray}
\begin{eqnarray}
H_4^{(64)}& = &
 -3  e^{2 \phi_0} \phi^{(2)}  
  \left(h_{12}-2 h_{32}\right)^{3}
%        \nonumber \\ & &
  +
  \frac{81}{5}  e^{2\phi_0} 
  \left(h_{12}-2 h_{32}\right)^2\left(h_{14c}+h_{14s}  \right)
%        \nonumber \\ & &
  -
  \frac{8}{3}  (\phi^{(2)})^{2}  
        \left(h_{12}-2  h_{32}\right)
        \nonumber \\ & &
  +
  \frac{2}{15}  
    \left(
    \left(
     10h_{12}\left(h_{12}-2  h_{32}\right) 
      +133  h_{14c}  
      +108  h_{14s}  
      +100  h_{34}  
    \right)
      \phi^{(2)}
  +110
    \left(h_{12}-2  h_{32}\right)
      \phi_{41}
    \right)
        \nonumber \\ & &
  +
  \frac{1}{720}
  \left(5536  a_{11}
    +648  a_{13}
    -20160  c_{32}
    +1725  h_{12}^{3}
    -3510  h_{12}^{2}  h_{32}
%    \right.
%        \nonumber \\ & &
%        \left.
    +4230  h_{12}  h_{14c}
    +6840  h_{12}  h_{34}
    -4500  h_{14c}  h_{32}
  \right)
\nonumber 
\end{eqnarray}
\begin{eqnarray}
H_4^{(62)}& = &
 \frac{3}{2}  e^{2 \phi_0}  \phi^{(2)}  
        \left(h_{12}-2 h_{32}\right)^{3}
%        \nonumber \\ & &
 - \frac{24}{5}  e^{2 \phi_0}
    \left(h_{12}-2 h_{32}\right)^2\left(h_{14c}+h_{14s}\right)
%        \nonumber \\ & &
  +
  \frac{4}{3}   (\phi^{(2)})^{2}  
        \left(h_{12} -2  h_{32} \right)
        \nonumber \\ & &
  +
  \frac{1 }{15}   
    \left(
    \left(
    5  h_{12}\left(h_{12} -2  h_{32} \right)
      -74  h_{14c}  
      -64  h_{14s}  
      -40  h_{34}  
    \right)
      \phi^{(2)}
      -80        
    \left(h_{12} -2  h_{32} \right)
      \phi_{41}
    \right)
        \nonumber \\ & &
  +
  \frac{1}{240}
  \left(
    -408  a_{11}
    -64  a_{13}
    +1920  c_{32}
    -115  h_{12}^{3}
    +240  h_{12}^{2}  h_{32}
%    \right.
%        \nonumber \\ & &
%        \left.
    -360  h_{12}  h_{14c}
    -480  h_{12}  h_{34}
    +420  h_{14c}  h_{32}
  \right)
\nonumber 
\end{eqnarray}

\subsection*{n = 4}

\begin{eqnarray} 
H_3^{(43)}& = &
 2      \phi^{(2)}  
  \left(h_{12}
    -2  h_{32}
  \right)
  +
  \frac{1}{8}
        \left(9  h_{12}^{2}
          -16  h_{12}  h_{32}
          -6  h_{14}
          +40  h_{34}
        \right)
\nonumber 
\end{eqnarray}
\begin{eqnarray} 
H_3^{(41)}& = &
 -2       \phi^{(2)}  
  \left(h_{12}
    -2  h_{32}
  \right)
  -
  \frac{1}{8}  
        \left(
          9  h_{12}^{2}
          -16  h_{12}  h_{32}
          -6  h_{14}
          +32  h_{34}
        \right)
\nonumber 
\end{eqnarray}
\begin{eqnarray} 
H_3^{(65)}& = &
 \frac{8}{3}  e^{2  \phi_0}
        \phi^{(2)}  
        \left(h_{12}-2 h_{32}\right)^{3}
  +
  \frac{4}{3}      (\phi^{(2)})^{2}  
        \left(h_{12}-2  h_{32} \right)
\nonumber \\ & &
  -
  \frac{1}{3}  
    \left(\left(5 \phi^{(2)} h_{12}+28 \phi_{41}\right)\left(h_{12}-2
    h_{32}\right)  
      -8  \phi^{(2)}\left( h_{14}  -4  h_{34} \right)  
    \right)
\nonumber \\ & &
  +
  \frac{1}{72}
    \left(-188  a_{12}
      +1512  a_{32}
      -147  h_{12}^{3}
      +297  h_{12}^{2}  h_{32}
      +333  h_{12}  h_{14}
      -576  h_{12}  h_{34}
      -342  h_{14}  h_{32}
    \right)
\nonumber 
\end{eqnarray}
\begin{eqnarray} 
H_3^{(63)}& = &
 -\frac{16}{3}  e^{2  \phi_0}  
         \phi^{(2)}  
        \left(h_{12}-2 h_{32}\right)^{3}
  -
  \frac{8}{3}     (\phi^{(2)})^{2}  
        \left(h_{12}-2  h_{32}
        \right)
\nonumber \\ & &
  +
  \frac{1}{3}   
    \left(\left(7  \phi^{(2)} h_{12} 
      +44  \phi_{41} \right)\left(h_{12}-2 h_{32}\right)
      -10\left( h_{14}  -4 h_{34}  \right)\phi^{(2)}
    \right)
\nonumber \\ & &
  +
  \frac{1}{72}
    \left(272  a_{12}
      -2016  a_{32}
      +219  h_{12}^{3}
      -441  h_{12}^{2}  h_{32}
      -486  h_{12}  h_{14}
      +864  h_{12}  h_{34}
      +486  h_{14}  h_{32}
    \right)
\nonumber 
\end{eqnarray}
\begin{eqnarray} 
H_3^{(61)}& = &
 \frac{8}{3}  e^{2  \phi_0}   
        \phi^{(2)}  
        \left(h_{12}-2 h_{32}\right)^{3}
  +
  \frac{4}{3}(\phi^{(2)})^{2}  
        \left(h_{12}-2  h_{32}\right)
\nonumber \\ & &
  -
  \frac{2    }{3} 
    \left(\left( \phi^{(2)}h_{12}
      +8  \phi_{41} \right)\left(h_{12}-2 h_{32}\right)
      -\left(h_{14}  
      -4  h_{34}\right)  \phi^{(2)}
    \right)
\nonumber \\ & &
  +
  \frac{1}{24} 
    \left(-28  a_{12}
      +192  a_{32}
      -24  h_{12}^{3}
      +48  h_{12}^{2}  h_{32}
      +51  h_{12}  h_{14}
      -96  h_{12}  h_{34}
      -48  h_{14}  h_{32}
    \right)
\nonumber 
\end{eqnarray}
\begin{eqnarray} 
H_4^{(42)}& = &
 -2     \phi^{(2)}  
  \left(h_{12}
    -2  h_{32}
  \right)
  -
  \frac{1}{8}\left(7  h_{12}^{2}
        -12  h_{12}  h_{32}
        -2  h_{14}
        +32  h_{34}\right)
\nonumber 
\end{eqnarray}
\begin{eqnarray} 
H_4^{(64)}& = &
 -\frac{16}{3}  e^{2  \phi_0} 
        \phi^{(2)}  
        \left(h_{12}-2 h_{32}\right)^{3}
  -
  \frac{8}{3}(\phi^{(2)})^{2}  
        \left(h_{12}-2  h_{32}\right)
\nonumber \\ & &
  +
  \frac{2    }{3}   
    \left(\left(2 \phi^{(2)}  h_{12}
      +22  \phi_{41}\right)\left(h_{12}-2 h_{32}\right) 
      -5 \left( h_{14}  
      -4  h_{34}\right)  \phi^{(2)}
    \right)
\nonumber \\ & &
  +
  \frac{1}{144}\left(416  a_{12}
    -4032  a_{32}
    +345  h_{12}^{3}
    -702  h_{12}^{2}  h_{32}
    -846  h_{12}  h_{14}
    +1368  h_{12}  h_{34}
    +900  h_{14}  h_{32}
  \right)
\nonumber 
\end{eqnarray}
\begin{eqnarray} 
H_4^{(62)}& = &
 \frac{8}{3}  e^{2  \phi_0}  \phi^{(2)}  
        \left(h_{12}-2 h_{32}\right)^{3}
  +
  \frac{4}{3}  (\phi^{(2)})^{2}  
        \left(h_{12}-2  h_{32}\right)
\nonumber \\ & &       
  +
  \frac{1}{3} 
    \left(\left( \phi^{(2)} h_{12}-16  \phi_{41}\right)\left(h_{12}-2
    h_{32}\right) 
      +2 \left( h_{14}  -4  h_{34}\right)  \phi^{(2)}
    \right)
\nonumber \\ & &       
  +
  \frac{1}{144}\left(-40  a_{12}
    +1152  a_{32}
    -69  h_{12}^{3}
    +144  h_{12}^{2}  h_{32}
    +216  h_{12}  h_{14}
    -288  h_{12}  h_{34}
    -252  h_{14}  h_{32}
  \right)
\nonumber 
\end{eqnarray}

\end{document}